\documentclass[11pt]{article} 
\usepackage[paperheight=9.44in,paperwidth=6.69in,left=1.85cm,right=1.8cm,bottom=2.5cm,top=3cm,twoside]{geometry}

\usepackage{subfig} 

\usepackage[ruled,vlined,boxed]{algorithm2e}
\usepackage{authblk} 
\usepackage{fancyhdr} 
\usepackage{lastpage} 
\usepackage{textcomp} 
\usepackage{amsmath} 
\usepackage{amssymb} 
\usepackage{amsthm} 
\usepackage{graphicx} 
\usepackage{multirow} 
\usepackage{hyperref} 
\usepackage{longtable} 
\usepackage{float} 

\newcommand{\linia}{\noindent\rule{\linewidth}{0.5mm}\hrulefill} 

\SetAlFnt{\small}
\SetAlCapFnt{\small}
\SetAlCapNameFnt{\small}
\SetAlCapHSkip{0pt}
\IncMargin{-\parindent}

\usepackage{times}
\usepackage{titlesec}

\titleformat*{\section}{\large\bfseries}
\titleformat*{\subsection}{\normalsize\bfseries}

\fancypagestyle{firststyle}
{
\fancyhf{}
\setcounter{page}{41} 
\fancyfoot{}
\fancyhead{}
\fancyhead[CE]{\upshape Zeszyty Naukowe WWSI, No 20, Vol. 13, 2019, pp.\thepage-\pageref{LastPage} \newline DOI: 10.26348/znwwsi.20.41}
\fancyhead[CO]{\upshape Zeszyty Naukowe WWSI, No 20, Vol. 13, 2019, pp. \thepage-\pageref{LastPage} \newline DOI: 10.26348/znwwsi.20.41}
\fancyfoot[L,R]{\footnotesize\bfseries Manuscript received April 23, 2019}
\fancyfoot[LE,RO]{\thepage}
}

\title{\large \bfseries Extension of the one-dimensional Stoney algorithm \\ to a two-dimensional case} 
\author{\normalsize Zenon Gniazdowski\thanks{E-mail: zgniazdowski@wwsi.edu.pl}}
\affil{\normalsize Warsaw School of Computer Science}

\date{\vspace{-5ex}}
\providecommand{\keywords}[1]{\textbf{\textit{Keywords ---}} #1}

\begin{document}

\maketitle 
\thispagestyle{firststyle} 

\linia
\begin{abstract}\label{abstract}
	\noindent This article presents the extension of the one-dimensional Stoney algorithm to a two-dimensional case. The proposed extension consists in modifying the method of curvature estimation. The surface profile of the wafer before deposition of the thin film and after its deposition was locally approximated by the quadric. From this quadric, a quadratic form and the first degree surface were separated. An eigenproblem was solved for the matrix of this quadratic form. From eigenvectors a new coordinate system was created in which a new formula of the quadric was found. In this new coordinate system, the two-dimensional problem of estimating the curvature tensor has been solved by solving two independent one-dimensional problems of curvature estimation. Returning to the primary coordinate system, in this primary system, a solution to the two-dimensional problem was obtained. The article proposes five versions of the two-dimensional Stoney algorithm, with diverse complexity and accuracy. The recommendation for the version of the algorithm that could be practically used was also presented.
\end{abstract}
\keywords{\small Stoney formula, Stoney equation, 2D Stoney algorithm, tensor of curvature, stress in a thin film, anisotropy, quadric, quadratic form}\label{keywords}

\section{Introduction}
Strong stress in a thin film deposited on the substrate may lead to degradation of the parameters of this substrate, and thus may prevent the formation of semiconductor devices on this substrate. Therefore, it is important to be able to estimate the level of this stress \cite{Piotrowskaetal.2006}.

The Stoney algorithm published in 1909 is used to identify stress in thin films \cite{Stoney1909}.
In Stoney's original work, the possibility of calculating the stress in a thin film of nickel applied to the surface of the steel ruler was shown.
The stresses in the nickel film were determined based on the estimate of the radius of curvature of the ruler after deposition of the film.
A number of simplifying assumptions have been adopted in the calculations \cite{Ardigo2014}:
\begin{itemize}
	\item It was assumed that the problem being solved is one-dimensional.
	\item Changes in the width of the ruler, described by the Poisson ratio, have not been taken into account.
	\item An anisotropy of the substrate was omitted.
	\item The initial curvature of the ruler that existed before the film was deposited is not taken into account.
	\item It was assumed that the thickness of the deposited film is much smaller than the thickness of the ruler.
\end{itemize}
On the one hand, the Stoney algorithm is used to calculate uniaxial stress that works in one dimension.
On the other hand, there is a need to estimate stress in thin films deposited on two-dimensional structures, such as semiconductor wafers.
This article will propose the extension of the one-dimensional Stoney algorithm to a form that, assuming the isotropy of the substrate, will allow to identify the two-dimensional stress in the deposited film.
The stress tensor estimated by Stoney equation is proportional to the change of curvature tensor.
The proportionality factor is the parameter related to material constants of the analyzed structure.
This parameter is not modified in this work. The modification of the Stoney algorithm presented in this article concerns only the modification of the method of estimating the curvature tensor, and in no way concerns the modification of the calculation of material parameters of the structure.
Therefore, in the experimental part of this article only the results of the estimation of change in the tensor of curvature will be shown and discussed, and the final estimation of the stress tensor will not be analyzed.

This new version of the algorithm will be presented in the form of tables with sub-algorithms containing a list of their successive steps.
On the right, in the form of analogous comments, such as those used in the C ++ programming language, references will be given, indicating previously defined formulas or tables with sub-algorithms.

At the end, anticipating possible misunderstandings related to the interpretation of terms or symbols  used in the description of some mathematical formulas, the intentions of the author of this article will be explained:
\begin{itemize}
	\item In the article, in the context of the complexity and accuracy of the algorithms, the word "ranking" will be used.
	      The meaning of the word will be the same as in sports competitions.
	      The most accurate algorithm will be in the first place in the accuracy ranking.
	      The most complex algorithm, in the ranking of complexity will be in the last place.
	\item In this article, the prime symbol $(')$ will be used to denote variables and functions referring to a particular coordinate system that is different from the primary coordinate system. In this article, this particular coordinate system will be the coordinate system related to the principal axes of a certain tensor. The only exception to this rule will be to describe the first derivative of the function in formula (\ref{Eq04}), using the prime symbol.
	\item The double prime symbol $({''})$ will be used in this article to denote the second derivative of the function of one variable. Only in the case of the second derivative of the function of two variables, or in the case of partial derivatives, the notation using the Jacobi "delta" $(\partial)$ will be used instead of the double prime symbol.
\end{itemize}

\section{Preliminaries}
To modify the Stoney algorithm, one should present basic definitions and equations, useful in further proceedings, leading to the identification of stress in the 2D structure.
\subsection{One-dimensional Stoney formula}
The stress in a thin film can be estimated from the following formula \cite{Ardigo2014}:
\begin{equation}\label{Eq01}
	\sigma_f=\alpha \cdot \Delta \kappa,
\end{equation}
where
\begin{equation}\label{Eq02}
	\alpha=\frac{E_s}{1-{\nu}_s} \cdot \frac{{t_s^2}}{6t_f}
\end{equation}
and
\begin{equation}\label{Eq03}
	\Delta \kappa = \kappa_b - \kappa_a = \frac{1}{r_b}-\frac{1}{r_a}.
\end{equation}
In formulas (\ref{Eq01}-\ref{Eq03}), the following designations were adopted:
\begin{itemize}
	\item $E_s$ - Young's modulus of the substrate,
	\item $\nu_s$ - Poisson's ratio of the substrate,
	\item $t_s$ and $t_f$ - the thickness of the substrate and deposited film, respectively,
	\item $\Delta\kappa$ - change of curvature resulting from deposition of the film,
	\item $r_a$ and $r_b$ - wafer curvature radii before and after the deposition process.
\end{itemize}
If at a given point the surface profile of the wafer is described by the function $f(x)$, then the radius of curvature in the formula (\ref{Eq03}) is determined by \cite{VanVliet1993}:
\begin{equation}\label{Eq04}
	r_x=\frac{(1+(f{'}(x))^2)^{\frac{3}{2}}}{f^{''}(x)}
\end{equation}
Stoney's formula (\ref{Eq01}) describes stress in a ruler, which appears as a result of depositing a thin film on its surface and describes only a one-dimensional case.
To describe a two-dimensional case, the above formula (\ref{Eq01}) should be extended and generalized.
\subsection{Description of anisotropy}
Anisotropy is based on the fact that certain properties or quantities (usually physical) are related to the direction. It is more precise to say that anisotropy exists when the observed quantities depend on the coordinate system. The opposite of anisotropy is isotropy. The tensor is the carrier of information about anisotropy. On the other hand, the scalar carries information about isotropy.

To define a tensor, one must assume the existence of a coordinate system. Physical quantities that are not dependent on this system are scalars. The quantities defined in the coordinate system are tensors. The rank of tensor manifests by the number of indexes in its description. In this way:
\begin{itemize}
	\item Scalar, which is the zero rank tensor, has no index in its description.
	\item The vector is the first rank tensor, so in its description there is only one index. For a given coordinate system, in the $n$-dimensional space the vector is completely defined by $n$ components, which are its orthogonal projections on the respective axes of the coordinate system.
	\item The second rank tensor has two indexes. It is a quantity that in the $n$-dimensional coordinate system is defined by $n \times n$ numbers that form a square matrix. An example of a second rank tensor is stress, strain as well as the matrix of quadratic form.
\end{itemize}
It should also be noted that:
\begin{itemize}
	\item There are higher rank tensors, but from the point of view of this article they are not important, so they will not be discussed.
	\item Second rank tensors are represented by square matrices, but not all square matrices are tensors.
\end{itemize}
\subsection{Tensor in the rotated coordinate system}
The rotation of the $OXY$ coordinate system is considered without changing its origin.
The axes of the coordinate system before rotation are described as $X$ and $Y$. The axes of the coordinate system after rotation are described as $X{'}$ and $Y{'}$.
\begin{table}
	\centering
	\caption{Direction cosines between axes before and after rotation}\label{tab1}
	\fontsize{10}{14}\selectfont{
		\begin{tabular}{c|c|c|c} \hline
			\multicolumn{2}{c|}{\ } & \multicolumn{2}{|c}{Axes before rotation}                                   \\ \cline{3-4}
			\multicolumn{2}{c|}{\ } & $X$                                       & $Y$                             \\ \hline \hline
			Axes after              & $X^{'}$                                   & $cos(X^{'},X)$ & $cos(X^{'},Y)$ \\ \cline{2-4}
			rotation                & $Y^{'}$                                   & $cos(Y^{'},X)$ & $cos(Y^{'},Y)$ \\ \hline
		\end{tabular}}
\end{table}
Table \ref{tab1} shows the directional cosines between the coordinate system axes after rotation and the coordinate system axes before  rotation \cite{Nye1957}.
The content of this table forms the rotation matrix $R$:
\begin{equation}\label{Eq05}
	R
		{=}
	\begin{bmatrix}
		cos(X^{'},X) & cos(X^{'},Y) \\
		cos(Y^{'},X) & cos(Y^{'},Y)
	\end{bmatrix}
	{=}
	\begin{bmatrix}
		r_{11} & r_{12} \\
		r_{21} & r_{22}
	\end{bmatrix}.
\end{equation}
The first index in $r_{ij}$ refers to the axis of the system after rotation. The second index refers to the axis before rotation.
Each row in the rotation matrix (\ref{Eq05}) represents two directional cosines of the $X{'}$ and $Y{'}$ axes relative to the orthogonal $X$ and $Y$ axes. Thus, the components of this matrix are mutually dependent, and the $R$ matrix is an orthogonal matrix:
\begin{equation}\label{Eq06}
	R^{-1} {=}R^T.
\end{equation}
If the set of tensor components before the rotation is known, and rotation (\ref{Eq05}) is known, then a set of components describing the tensor in the new coordinate system can be found.
Table \ref{tab2} shows the transformation laws from zero rank tensors up to tensors rank two \cite{Nye1957}.

\begin{table}
	\centering
	\caption{Description of tensor transformation}\label{tab2}
	\fontsize{10}{14}\selectfont{
		\begin{tabular}{c|c|c|c} \hline
			Rank               & New components                & Old components                & \multirow{2}{*}{Note}            \\
			of	tensor           & expressed by old              & expressed by new              &                                  \\ \hline \hline
			\multirow{2}{*}{0} & \multirow{2}{*}{$g '=g$}      & \multirow{2}{*}{$g =g '$}     & $g$, $g'$ $-$ scalar             \\
			                   &                               &                               & (zero rank tensor)               \\ \hline
			\multirow{2}{*}{1} & \multirow{2}{*}{$v'=R v$}     & \multirow{2}{*}{$v=R^T v'$}   & {$v$, $v'$ $-$ vector}           \\
			                   &                               &                               & (first rank tensor)              \\ \hline
			\multirow{2}{*}{2} & \multirow{2}{*}{$M'=R M R^T$} & \multirow{2}{*}{$M=R^T M' R$} & $M$, $M'$ $-$ second rank tensor \\
			                   &                               &                               & expressed as a square matrix     \\ \hline
		\end{tabular}}
\end{table}
\subsection{Quadric}
The Cartesian coordinate system $OXYZ$ is given.
The axes of this system have directions in accordance with the directions of unit vectors: $X=[1,0,0]^T$, $Y=[0,1,0]^T$ and $Z=[0,0,1]^T$.
In this space, the quadratic function of two variables $z=f(x,y)$ is considered:
\begin{equation}\label{Eq07}
	z=f(x,y)=ax^2+by^2+2cxy+dx+ey+g.
\end{equation}
This function is a special case of a quadric \cite{Groshong1989} \cite{Bronsztejn2004}, which describes the surface hanging above  the $OXY$ plane. For further analysis, the following denotations are used:
\begin{equation}\label{Eq08}
	p=[x,y]^{T},
\end{equation}
\begin{equation}\label{Eq09}
	M {=}
	\begin{bmatrix}
		a & c \\
		c & b\end{bmatrix},
\end{equation}
\begin{equation}\label{Eq10}
	v=[d,e]^T.
\end{equation}
The components of matrix $M$ and vector $v$ are coefficients that appear in expression (\ref{Eq07}).
The vector $p$ (\ref{Eq08}) represents a point on the $OXY$ plane.
Using the notation (\ref{Eq08}) - (\ref{Eq10}), the surface (\ref{Eq07}) can be represented in the form of the following sum:
\begin{equation}\label{Eq11}
	f(p)=p^{T}Mp+p^{T}v+g.
\end{equation}
The above sum represents the superposition of three different surfaces.
The first surface is a quadratic form: $q(p)=p^{T}Mp$.
The second surface is the plane $l(p)=p^{T}v$.
The last surface is the plane $g$ parallel to the $OXY$ plane:
\begin{equation}\label{Eq12}
	f(p)=q(p)+l(p)+g.
\end{equation}
It should be noted that each of the three separate surfaces is represented by the corresponding tensor.
The quadratic form $q(p)$ is represented by a symmetric matrix (\ref{Eq09}) of a quadratic form, which is the second rank tensor:
\begin{equation}\label{Eq13}
	q(p)=p^{T}Mp=ax^{2}+by^{2}+2cxy.
\end{equation}
The plane $l(p)$ is represented by the vector $v$ (\ref{Eq10}) normal to it, being the first rank tensor:
\begin{equation}\label{Eq14}
	l(p)=p^{T}v=dx+ey.
\end{equation}
The surface parallel to the $OXY$ plane represents the scalar $g$, which is the zero rank tensor.

\begin{table}
	\centering
	\caption{Example of a set of points describing the surface profile of a wafer}\label{tab3}
	\fontsize{10}{14}\selectfont{
		\begin{tabular}{c|c|c} \hline
			$x$      & $y$      & $z$      \\ \hline \hline
			$x_0$    & $y_0$    & $z_0$    \\ \hline
			$x_1$    & $y_1$    & $z_1$    \\ \hline
			$x_2$    & $y_2$    & $z_2$    \\ \hline
			$\ldots$ & $\ldots$ & $\ldots$ \\ \hline
			$x_n$    & $y_n$    & $z_n$    \\ \hline
		\end{tabular}}
\end{table}

\subsection{Identification of the quadric}
\begin{figure}
	\centering
	\includegraphics[width=7cm]{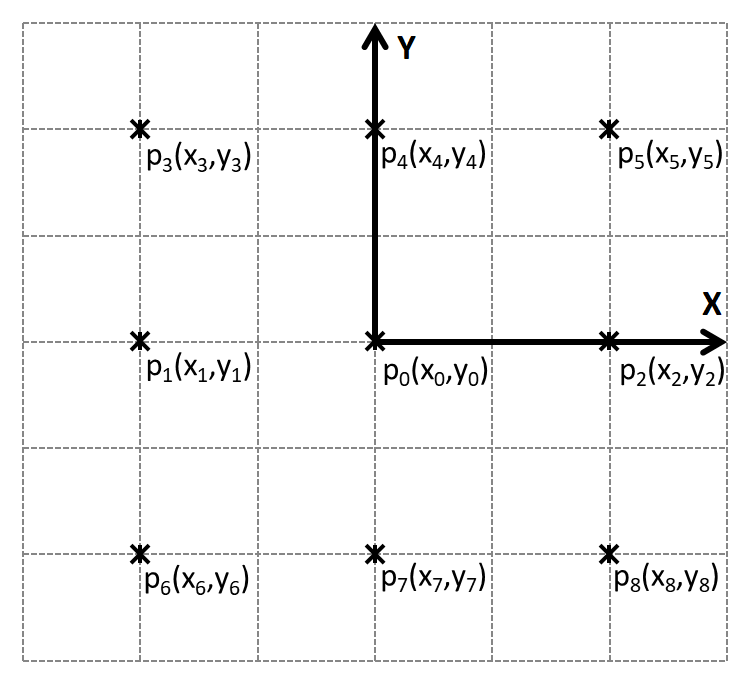}
	\caption{An example of point $p_0$ together with its neighbors}\label{fig1}
\end{figure}

In the three-dimensional space a set of $p_i$ points, described by three numbers $(x_i,y_i,z_i)$ is given.
It is known that these points are on a surface $z=f(x,y)$.
Despite the fact that the surface formula is unknown, it is not unreasonable to assume that near the given point this function can be well approximated by the quadric (\ref{Eq07}).
Selecting on the $OXY$ plane a point described by a pair of numbers marked as $ (x_0,y_0) $ together with the corresponding value $z_0$, and in a similar way by selecting a few of its nearest neighbors, the quadric (\ref{Eq07}) can be identified by the least squares method.
Table \ref{tab3} shows an example of a set of points describing a fragment of the surface profile of a wafer being measured.
Figure \ref{fig1} presents an example of the distribution of points needed to identify the quadric around point $p_0$.
Neighbors of point $p_0$ are points $p_1,p_2,...,p_8$.
It has been assumed that point $p_0$ will be the center $OXY$ of the new coordinate system in which the quadric will be identified.
Three numbers $(x_0,y_0,z_0)$ in the new coordinate system will become a three numbers $(0,0,z_0)$.
Similarly, all sets of three numbers that describe the nearest neighbors of point $p_0$ will be described in this new $OXY$ system.
Now the three numbers $(x_i,y_i,z_i)$ will become the three numbers $(x_i-x_0,y_i-y_0,z_i)$.

Identification of the quadric (\ref{Eq07}) consists in finding unknown set of coefficients $\beta$ in equation (\ref{Eq07}):
\begin{equation}\label{Eq15}
	\beta=[a,b,2c,d,e,g]^T.
\end{equation}
Vector $\beta$ (\ref{Eq15}) can be identified by the least squares method.
For this purpose, first of all, components of the system of normal equations should be found, and then this system should be solved \cite{Groshong1989}\cite{Manczak1971}.
First, an $X$ matrix is formed based on Table \ref{tab3}:
\begin{equation}\label{Eq16}
	X {=}
	\begin{bmatrix}
		x_0^2  & y_0^2  & x_0y_0 & x_0    & y_0    & 1      \\
		x_1^2  & y_1^2  & x_1y_1 & x_1    & y_1    & 1      \\
		\vdots & \vdots & \vdots & \vdots & \vdots & \vdots \\
		x_n^2  & y_n^2  & x_ny_n & x_n    & y_n    & 1      \\
	\end{bmatrix}.
\end{equation}
The matrix $A$ will be formed from the matrix $X$ (\ref{Eq16}), which is the matrix of the system of normal equations:
\begin{equation}\label{Eq17}
	A=X^TX.
\end{equation}
Next, using the vector $z=[z_0,z_1,\ldots z_n]^T$ (see Table \ref{tab3}) and the matrix $X$ (\ref{Eq16}), vector $\gamma$ of the right side of the system of normal equations is formed:
\begin{equation}\label{Eq18}
	\gamma=X^Tz.
\end{equation}
In the next step, the system of normal equations is solved:
\begin{equation}\label{Eq19}
	A\beta=\gamma.
\end{equation}
Identified coefficients $\beta$ (\ref{Eq15}) will be used to form both the matrix (\ref{Eq09}), which describes the quadratic form (\ref{Eq13}), as well as the vector (\ref{Eq10}), which describes the plane (\ref{Eq14}).
A detailed algorithm for identifying a vector containing the quadric coefficients (\ref{Eq15}) is shown in Table \ref{findQuadric}.

\begin{table}
	\centering
	\caption{The algorithm for finding quadric}\label{findQuadric}
	\fontsize{10}{14}\selectfont{
		\begin{tabular}{rll} \hline
			\textbf{Input}:  & The surface profile $z(x,y)$ around point $p_0$                 & //Tab. \ref{tab3} \\
			\textbf{Output}: & A set $\beta$ containing the coefficients of quadric $z=f(x,y)$ & //(\ref{Eq15})    \\ \hline \hline
			\textbf{Begin}   &                                                                                     \\
			1.               & Form the matrix $X$;                                            & //(\ref{Eq16})    \\
			2.               & Create matrix $A$;                                              & //(\ref{Eq17})    \\
			3.               & Find $\gamma$;                                                  & //(\ref{Eq18})    \\
			4.               & Solve the system of normal equations;                           & //(\ref{Eq19})    \\
			\textbf{End}.    &                                                                                     \\ \hline
		\end{tabular}}
\end{table}
Finally, it should be noted that the number of identified coefficients of quadric imposes a condition on the number of rows in the matrix (\ref{Eq16}), and thus imposes a condition on the minimum number of neighbors of point $p_0$, which are necessary to identify the quadric around point $p_0$.
In order to be able to identify the quadric, the number of nearest neighbors of point $p_0$ together with point $p_0$, can not be smaller than the number of coefficients in the formula of quadric. This means that the number of rows in the matrix (\ref{Eq16}) can not be smaller than the number of columns in this matrix.
\subsection{Quadric in the rotated coordinate system}
Suppose that in the center of the $OXY$ coordinate system is an observer who sees the surface (\ref{Eq07}) hanging above him.
The observer does not change his position, but only rotates in the $OXY$ plane, around an axis $Z$ perpendicular to the $OXY$ plane.
Let the axes of the coordinate system rotate along with the observer.
As a result, the $X$ axis will become the $X{'}$ axis, and the $Y$ axis will become the $Y{'}$ axis.
Rotating the coordinate system does not change the surface, but it will change its perception in the new coordinate system.
The observer may conclude that the observed surface is anisotropic.
The challenge for him is to find a way to describe this surface in any rotated coordinate system other than the initial coordinate system.

The surface (\ref{Eq07}) seen by the observer is a superposition of three surfaces (\ref{Eq11}):
\begin{itemize}
	\item a quadratic form (\ref{Eq13}), which is described by the matrix $M$ (\ref{Eq09}),
	\item a plane (\ref{Eq14}) that the vector $v$ (\ref{Eq10}) describes,
	\item a plane parallel to the plane $OXY$, defined by the constant $g$.
\end{itemize}
It is obvious that the rotation will not change the constant $g$.
The problem of finding the equation of the observed surface in the new coordinate system is simple (or complex), as simple (or complex) is to find the first rank tensor (vector) or the second rank tensor (quadratic form matrix) in the new coordinate system.
For this purpose, it is necessary to know both coordinate systems: the initial coordinate system (before rotation) and the final coordinate system (after rotation).

\subsubsection{Principal axes of the quadratic form}\label{Subsubsection:Principal axes of the quadratic form}
The matrix $M$ (\ref{Eq09}), which describes the quadratic form in the old coordinate system, will be represented in the new coordinate system by the new matrix $M{'}$.
An observer who wants to determine the curvature at some point $p_0$ would like to find a coordinate system in which the derivative of the function describing the observed surface, calculated with respect to one variable, will not depend on the value of the second variable.
This is possible only when the equation of the quadratic form will be known in the system of its principal axes.
This means that the matrix of the quadratic form $M$ should be diagonalized by solving the eigenproblem.

The $M$ matrix is the second rank tensor.
By solving the eigenproblem for matrix $M$, we will obtain eigenvalues $\lambda_1$ and $\lambda_2$, as well as their corresponding eigenvectors $u_1=[u_{11},u_{12}]^T$ and $u_2=[u_{21},u_{22}]^T$.
The eigenvectors after normalization to the unit length determine the directions of the new axes of the orthogonal coordinate system $OX{'}Y{'}$.
They are so-called principal axes (or principal directions) of the quadratic form.
In the system of this principal axes, matrix $M$ becomes a diagonal matrix $M{'}$:
\begin{equation}\label{Eq20}
	M{'} {=}
	\begin{bmatrix}
		\lambda_1 & 0         \\
		0         & \lambda_2\end{bmatrix}.
\end{equation}
At the same time, the eigenvectors $u_1$ and $u_2$ will be successive rows of the orthogonal matrix $R$ (\ref{Eq05}), describing the transformation matrix $M$ (\ref{Eq09}) to the diagonal form $M{'}$ (\ref{Eq20}) \cite{Gniazdowski2017}:
\begin{equation}\label{Eq21}
	R {=}
	\begin{bmatrix}
		u_{11} & u_{12} \\
		u_{21} & u_{22}\end{bmatrix}.
\end{equation}
This transformation (see Table \ref{tab2}) takes the following form:
\begin{equation}\label{Eq22}
	M{'}=RMR^T.
\end{equation}
In a similar way,
the matrix $R$ also allows to return from the form of a diagonal matrix $M{'}$ to the primary form $M$:
\begin{equation}\label{Eq23}
	M=R^TM{'}R.
\end{equation}
\subsubsection{Quadric in the coordinate system of the principal axes of its quadratic form}
The transition to the coordinate system described by the principal axes of a quadratic form is equivalent to the rotation of the primary $OXY$ coordinate system to the final coordinate system $OX{'}Y{'}$.
The equation of the quadratic form $q(x,y)$, which in the $OXY$ coordinate system was described by equation (\ref{Eq13}), in the new coordinate system $OX{'}Y{'}$ is described by the formula $q{'}(x{'},y{'})$:
\begin{equation}\label{Eq24}
	q{'}(x{'},y{'})=\lambda_1 \cdot (x{'})^2+\lambda_2 \cdot (y{'})^2.
\end{equation}
On the other hand, the plane $l(x,y)$ having the form (\ref{Eq14}) is represented by the vector $v$ (\ref{Eq10}), perpendicular to it.
Rotation of the coordinate system will not change the plane, and therefore also the vector $v$, which represents it.
However, in the new coordinate system the mathematical description of the vector will change.
This vector will now be represented by new components (see Table 2):
\begin{equation}\label{Eq25}
	v{'}=Rv.
\end{equation}
This means that with the rotation of the coordinate system, the vector $v=[d,e]^T$ will become the vector $v{'}=[d{'},e{'}]^T$.
Therefore, in the system of a new $X{'}$ and $Y{'}$ axes, the plane equation will be in the form $l{'}(x{'},x{'})$:
\begin{equation}\label{Eq26}
	l{'}(x{'},y{'})=d{'}x{'}+e{'}y{'}.
\end{equation}
In this way, the vector $\beta$ (\ref{Eq15}) will become the vector $\beta{'}$:
\begin{equation}\label{Eq27}
	\beta{'}=[\lambda_1,\lambda_2,0,d{'},e{'},g]^T.\end{equation}
Finally, the surface, which in the primary coordinate system is described by function (\ref{Eq07}), after transformation to the coordinate system defined by the principal axes of a quadratic form, will take the following form:
\begin{equation}\label{Eq28}
	z=f{'}(x{'},y{'})=\lambda_1 \cdot (x{'})^2+\lambda_2 \cdot (y{'})^2+d{'}x{'}+e{'}y{'}+g.
\end{equation}
The observer has achieved his goal defined in subsection \ref{Subsubsection:Principal axes of the quadratic form}.
He managed to find a coordinate system in which the derivative of the function describing the surface to be seen, calculated with respect to one variable, does not depend on the value of the second variable. An  algorithm for finding a quadric in a coordinate system formed by the principal axes of its quadratic form is shown in Table \ref{quadricInPrincipal}.

\begin{table}
	\centering
	\caption{The algorithm of finding a quadric in a coordinate system formed from the principal axes of its quadratic form}\label{quadricInPrincipal}
	\fontsize{10}{14}\selectfont{
		\begin{tabular}{rll} \hline
			\textbf{Input}:  & Vector $\beta$ containing the coefficients of quadric                                        & //(\ref{Eq15}) \\
			\textbf{Output}: & Vector $\beta{'}$ in the system of the principal axes of a quadratic form                    & //(\ref{Eq27}) \\ \hline \hline
			\textbf{Begin}   &                                                                                              &                \\
			1.               & From vector $\beta$ (\ref{Eq15}), create matrix $M$;                                         & //(\ref{Eq09}) \\
			2.               & From vector $\beta$ (\ref{Eq15}), create vector $v$;                                         & //(\ref{Eq10}) \\
			3.               & Solve eigenproblem for matrix $M$ (\ref{Eq09});                                              &                \\
			4.               & Create matrix $M{'}$ from eigenvalues;                                                       & //(\ref{Eq20}) \\
			5.               & Create matrix $R$ from eigenvectors;                                                         & //(\ref{Eq21}) \\
			6.               & Find vector $v{'}$;                                                                          & //(\ref{Eq25}) \\
			7.               & Using matrix $M{'}$ (\ref{Eq20}) and vector $v{'}$ (\ref{Eq25}), form the vector $\beta{'}$, & //(\ref{Eq27}) \\
			                 & which contains the coefficients of the quadric (\ref{Eq28})                                  &                \\
			                 & in the coordinate system of principal axes of its quadratic form (\ref{Eq13});               &                \\
			\textbf{End}.    &                                                                                              &                \\ \hline
		\end{tabular}}
\end{table}

\subsection{An algorithm for estimating the local curvature at a given point $p_0$}\label{Subsection:localCurvature}
In one-dimensional case, the curvature is calculated as the inverse of the radius described by the formula (\ref{Eq04}).
However, the use of this formula for a two-dimensional structure induces an additional difficulty.
The first derivative of the quadric (\ref{Eq07}), calculated with respect to one variable, is also a function of the second variable.
This means that curvature in a given direction is not only a function of this direction.
\begin{table}
	\centering
	\caption{The algorithm for identifying the tensor of curvature at point $p_0$}\label{findCurvatureTensor}
	\fontsize{10}{14}\selectfont{
		\begin{tabular}{rll} \hline
			\textbf{Input}:  & Vector $\beta{'}$ in the system of the principal axes of a quadratic form              & //(\ref{Eq27}) \\
			\textbf{Output}: & Tensor of curvature in the primary coordinate system                                   & //(\ref{Eq31}) \\ \hline \hline
			\textbf{Begin}   &                                                                                                         \\
			1.               & Find the principal curvatures in point $p_0$:                                          &                \\
			                 & (a) By differentiating (\ref{Eq28}) with respect to $x$, estimate the radius $r_x{'}$; & //(\ref{Eq04}) \\
			                 & (b) By differentiating (\ref{Eq28}) with respect to $y$, estimate the radius $r_y{'}$; & //(\ref{Eq04}) \\
			2.               & Find the diagonal curvature tensor in the system of its principal axes:                & //(\ref{Eq30}) \\
			                 & (a) $\kappa_{11} = 1 / r_x{'}$;                                                        &                \\
			                 & (b) $\kappa_{22} = 1 / r_y{'}$;                                                        &                \\
			3.               & Estimate the curvature tensor $\kappa$ in the primary coordinate system;               & //(\ref{Eq31}) \\
			\textbf{End}.    &                                                                                        &                \\ \hline
		\end{tabular}}
\end{table}

The analysis of formulas (\ref{Eq07}) and (\ref{Eq28}) shows that the above problem will not appear when the corresponding derivative will be calculated in the system of the principal axes of a quadratic form (\ref{Eq13}).
More specifically, in the system of the principal axes of a quadratic form, two independent principal curvatures can be estimated which will allow to identify the diagonal tensor of the curvature $\kappa{'}$ in the system of its principal axes:
\begin{equation}\label{Eq30}
	\kappa{'} {=}
	\begin{bmatrix}
		\kappa_{11} & 0           \\
		0           & \kappa_{22}\end{bmatrix}.
\end{equation}
The principal axes (eigenvectors) form the rotation matrix $R$ (\ref{Eq21}).
With this matrix, you can easily return to the primary coordinate system.
For this purpose, it is enough to perform an analogous transformation like the transformation in Table \ref{tab2} for the second rank tensor:
\begin{equation}\label{Eq31}
	\kappa=R^T\kappa{'}R
\end{equation}
The proposed approach allows to find a solution of a two-dimensional problem by solving two independent one-dimensional problems in a particular coordinate system.
Having obtained two independent one-dimensional solutions, you can find a solution of the two-dimensional problem in the primary coordinate system.
Finally, three detailed algorithms are used to identify the current curvature at point $p_0$.
The first is the algorithm described in Table \ref{findQuadric}, allowing the identification of the quadric.
The second one is the algorithm of finding a quadric in the coordinate system determined by the principal axes of its quadriatc form, presented in Table \ref{quadricInPrincipal}.
The last is the algorithm for estimating the curvature tensor at point $p_0$, described in Table \ref{findCurvatureTensor}.
\section{Two-dimensional Stoney algorithm}
It is assumed that the stress estimated with equation (\ref{Eq01}) works along the ruler and is a uniaxial stress.
Formula (\ref{Eq01}) shows this stress as the product of $\alpha$ and $\Delta\kappa$ components.
The $\alpha$ component contains the material constants of the substrate as well as the thickness of both the substrate and deposited film, in the analyzed structure.
Assuming that the wafer material is isotropic, the $\alpha$ component is a scalar that, regardless of whether the stress is analyzed in a one-dimensional ruler or in a two-dimensional wafer, can be described by formula (\ref{Eq02}).

On the other hand, the parameter $ \Delta \kappa$ (\ref{Eq03}) represents the deformation of the substrate described by the change in its curvature, which is caused by the stress resulting from depositing a thin film on the substrate.
In the one-dimensional structure of the ruler, curvature $\kappa$ is a number.
This means that in the one-dimensional case the stress acts along the ruler and has the character of uniaxial stress.
In the two-dimensional structure of the wafer, the curvature $\kappa$ is the second rank tensor, but also the stress is the second rank tensor.
To account for the two-dimensional nature of stress, the calculation of  curvature change (\ref{Eq03}) should be modified in the Stoney algorithm.

\begin{table}
	\centering
	\caption{The algorithm for finding stress in point $p_0$}\label{tabMainStress}
	\fontsize{10}{14}\selectfont{
		\begin{tabular}{rll} \hline
			\textbf{Input}:  & (a) A local profile $z_a(x,y)$ before the film was deposited               & //Tab. \ref{tab3} \\
			                 & (b)  A local profile $z_b(x,y)$  after the film was deposited              & //Tab. \ref{tab3} \\
			                 & (c) $E_s$ - Young's modulus of the substrate                               &                   \\
			                 & (d) $\nu_s$ - Poisson's ratio of the substrate                             &                   \\
			                 & (e) $t_s$ - thickness of the substrate                                     &                   \\
			                 & (f) $t_f$ - thickness of the deposited film                                &                   \\
			\textbf{Output}: & Stress tensor in point $p_0$                                               &                   \\ \hline \hline
			\textbf{Begin}   &                                                                            &                   \\
			1.               & Calculate material parameter $\alpha$ ;                                    & //(\ref{Eq02})    \\
			2.               & Estimate the change of curvature tensor $\Delta\kappa$;                    & //(\ref{Eq29})    \\
			3.               & Find the product of $\alpha$ (\ref{Eq02}) and $\Delta\kappa$ (\ref{Eq29}); & //(\ref{Eq01})    \\
			\textbf{End}.    &                                                                            &                   \\ \hline
		\end{tabular}}
\end{table}

While maintaining the assumption about the substrate isotropy, it is not necessary to change the way of calculating the $\alpha$ parameter.
This means that at a given value of parameter $\alpha$, any changes in stress will depend on the difference in curvature of the wafer after the film is deposited on the wafer and before its deposition.
However, now the difference in curvatures (\ref{Eq03}) will be replaced by the difference in tensors of curvature:
\begin{equation}\label{Eq29}
	\Delta \kappa=\kappa_b-\kappa_a.
\end{equation}
An algorithm for finding stress at point $p_0$ is presented in Table. {\ref{tabMainStress}. This algorithm is almost identical to the one-dimensional Stoney algorithm (\ref{Eq01}). The only difference between the algorithm described here and the one-dimensional Stoney algorithm is that the curvature is estimated differently.
The algorithm proposed in this article takes into account the fact that the problem being solved is a two-dimensional problem. This means that the curvature is the second rank tensor. The proposed changes are discussed in detail in the following sections of this article.

\subsection{Two-dimensional algorithm for estimating the change in the curvature tensor - Algorithm A}
\begin{table}
	\centering
	\caption{Algorithm A - The main algorithm used to identify the change in the tensor of curvature - a two-dimensional extension of the classic Stoney algorithm}\label{AlgorithmA}
	\fontsize{10}{14}\selectfont{
		\begin{tabular}{rll} \hline
			\textbf{Input}:  & (a) A local profile $z_a(x,y)$ before the film was deposited               & //Tab. \ref{tab3}               \\
			                 & (b)  A local profile $z_b(x,y)$  after the film was deposited              & //Tab. \ref{tab3}               \\
			\textbf{Output}: & Change of curvature tensor $\Delta\kappa$ in the primary coordinate system & //(\ref{Eq29})                  \\ \hline \hline
			\textbf{Begin}   &                                                                            &                                 \\
			1.               & Identify the quadric $z_a=f_a(x,y)$;                                       & //Tab. \ref{findQuadric}        \\
			2.               & Find the above quadric in the principal axes of its quadratic form;        & //Tab. \ref{quadricInPrincipal} \\
			3.               & At point $p_0$, find principal components of the curvature tensor change:  & //(\ref{Eq36})                  \\
			                 & (a) Estimate $\kappa{'}_{a11}$  with respect to $x^{'}$;                   & //(\ref{Eq04})                  \\
			                 & (b) Estimate $\kappa{'}_{a22}$ with respect to $y^{'}$;                    & //(\ref{Eq04})                  \\
			4.               & Estimate the tensor  $\kappa_a$ in the primary coordinate system;          & //(\ref{Eq31})                  \\
			5.               & Identify the quadric $z_b=f_b(x,y)$;                                       & //Tab. \ref{findQuadric}        \\
			6.               & Find the above quadric in the principal axes of its quadratic form;        & //Tab. \ref{quadricInPrincipal} \\
			7.               & At point $p_0$, find principal components of the curvature tensor change:  & //(\ref{Eq36})                  \\
			                 & (a) Estimate $\kappa{'}_{b11}$  with respect to $x^{'}$;                   & //(\ref{Eq04})                  \\
			                 & (b) Estimate $\kappa{'}_{b22}$ with respect to $y^{'}$;                    & //(\ref{Eq04})                  \\
			8.               & Estimate the tensor  $\kappa_b$ in the primary coordinate system;          & //(\ref{Eq31})                  \\
			9.               & Estimate the change of curvature tensor $\Delta\kappa$;                    & //(\ref{Eq29})                  \\
			\textbf{End}.    &                                                                            &                                 \\ \hline
		\end{tabular}}
\end{table}

Finding the difference (\ref{Eq29}) can be separated into two independent steps.
In the first stage, the curvature of the wafer profile ($\kappa_a$) should be found before the film is deposited on the wafer. In the second stage you have to find the curvature of the wafer profile ($\kappa_b$) after deposition the film on the wafer.
At the end of this procedure, the difference (\ref{Eq29}) of both curvatures  should be estimated.

Finding the curvature at a given moment (before deposition the film on the wafer and after its deposition) consists of several steps.
In the first step, you should identify the quadric describing the current wafer profile. Next, this quadric has to be described in the coordinate system of the principal axes of its quadratic form. In this system, the principal curvatures of the wafer profile, which will form a diagonal tensor of curvature (\ref{Eq30}), should be estimated.
Using the transformation (\ref{Eq31}), this tensor should be transformed into the primary coordinate system.
Having the tensors of the wafer profile curvatures before the film is deposited on the wafer and after its deposition, one can proceed to estimate the tensor difference between the two curvatures.
A detailed algorithm for identifying the difference between the two tensors is presented in Table \ref{AlgorithmA}.

\section{Simplification of the change of curvature estimation algorithm}
If the condition is met:
\begin{equation}\label{Eq31a}
	\left| f'(x)\right|  << 1,
\end{equation}
then the expression (\ref{Eq04}) can be simplified:
\begin{equation}\label{Eq32}
	r_x=\frac{1}{f_x^{''}(x)}.
\end{equation}
In this case, the curvature for a one-dimensional structure, as well as the principal curvature towards a certain $X{'}$ axis, in the coordinate system formed from the principal axes, is equal to:
\begin{equation}\label{Eq33}
	\kappa_x=\frac{1}{r_x}=f_x^{''}(x).
\end{equation}
Thanks to this, the formula (\ref{Eq29}) can also be simplified:
\begin{equation}\label{Eq34}
	\Delta \kappa=\kappa_b-\kappa_a=f_b^{''}(x,y)-f_a^{''}(x,y).
\end{equation}
Since the difference of the second derivatives of the two functions is equal to the second derivative of the difference of these functions, the expression (\ref{Eq34}) can be further simplified.
The above simplifications suggest several possible ways to further changes the algorithm described in chapter \ref{Subsection:localCurvature}, which will be presented below.

\begin{table}
	\centering
	\caption{Algorithm  B1 - The first simplification of the algorithm to identify change in curvature tensor - uses the difference in wafer profile models after deposition of a thin film and before its deposition}\label{AlgorithmB1}
	\fontsize{10}{14}\selectfont{
		\begin{tabular}{rll} \hline
			\textbf{Input}:  & (a) A local profile $z_a(x,y)$ before the film was deposited                                          & //Tab. \ref{tab3}               \\
			                 & (b)  A local profile $z_b(x,y)$  after the film was deposited                                         & //Tab. \ref{tab3}               \\
			\textbf{Output}: & Change of curvature tensor $\Delta\kappa$ in the primary coordinate system                            & //(\ref{Eq31})                  \\ \hline \hline
			\textbf{Begin}   &                                                                                                                                         \\
			1.               & Identify the quadric $z_a=f_a(x,y)$ before deposition the film;                                       & //Tab. \ref{findQuadric}        \\
			2.               & Identify the quadric $z_b=f_b(x,y)$ after deposition the film;                                        & //Tab. \ref{findQuadric}        \\
			3.               & Find the quadric $f_{b-a}$ as the difference of the two above quadrics;                               & //(\ref{Eq35})                  \\
			4.               & Find the above quadric in the principal axes of its quadratic form;                                   & //Tab. \ref{quadricInPrincipal} \\
			5.               & At point $p_0$, find principal components of the curvature tensor change:                             & //(\ref{Eq36})                  \\
			                 & (a) Estimate $\Delta\kappa{'}_{11}$ as the second derivative of (\ref{Eq35}) with respect to $x^{'}$; &                                 \\
			                 & (b) Estimate $\Delta\kappa{'}_{22}$ as the second derivative of (\ref{Eq35}) with respect to $y^{'}$; &                                 \\
			6.               & Estimate the tensor  $\Delta\kappa$ in the primary coordinate system;                                 & //(\ref{Eq31})                  \\
			\textbf{End}.    &                                                                                                       &                                 \\ \hline
		\end{tabular}}
\end{table}

\subsection{The first simplification - Algorithm  B1}
To make the first simplifying change of the algorithm from section \ref{Subsection:localCurvature}, the difference of functions describing the surface profiles of the wafer after deposition of the thin film on it and before its deposition will be described as $f_{b-a}(x,y)$:
\begin{equation}\label{Eq35}
	f_{b-a}(x,y)=f_b(x,y)-f_a(x,y).
\end{equation}
The change of  $\Delta\kappa$ can now be estimated as the second derivative of the difference (\ref{Eq35}).
After transforming the quadric (\ref{Eq35}) into the form $f^{'}_{b-a}$ described in the coordinate system formed by the principal axes of its quadratic form (see algorithm in Table \ref{quadricInPrincipal}), change of curvature $\Delta\kappa{'}$ can be expressed as follows:
\begin{equation}\label{Eq36}
	\Delta\kappa{'} = \begin{bmatrix}
		\frac{\partial{^{2}f^{'}_{b-a}}}{\partial{x^{'2}}} & 0                                                  \\
		0                                                  & \frac{\partial{^{2}f^{'}_{b-a}}}{\partial{y^{'2}}}
	\end{bmatrix}.
\end{equation}
After finding the change of curvature tensor in the system of the principal axes, one should return to the primary coordinate system.
A detailed algorithm for finding this change of curvature is presented in Table \ref{AlgorithmB1}.

\subsection{Second simplification - Algorithm  B2}
The next modification of the algorithm described in section \ref{Subsection:localCurvature} has a slightly different character.
In the algorithm shown in Table \ref{AlgorithmB1}, the change of curvature was calculated based on the second derivative of the difference between the two quadrics identified after and  before the deposition of the thin film.
In this context, the question arises whether the change of curvature of the wafer surface profile can be estimated using the quadric identified for the difference in wafer surface profiles after deposition the thin film and before it is deposited:
\begin{equation}\label{Eq37}
	\Delta_z(x,y)=z_b(x,y)-z_a(x,y).
\end{equation}
If this possibility exists, then the alternative algorithm should first find a quadric describing this difference:
\begin{equation}\label{Eq38}
	\Delta_z(x,y)=f(x,y).
\end{equation}
Next, this quadric should be transformed into a form whose description will refer to the coordinate system formed from the principal axes of its quadratic form.
Now, the change of curvature tensor in the coordinate system created by the principal axes can be estimated.
The obtained tensor is described by a diagonal matrix, which should eventually be transformed into the primary coordinate system.
In Table \ref{AlgorithmB2}, this version of the algorithm is also presented.
\begin{table}
	\centering
	\caption{Algorithm B2 - The second simplification of the algorithm to identify the change in the tensor of curvature - uses a model describing the difference in the profile of the wafer after deposition of the thin film and before its deposition}\label{AlgorithmB2}
	\fontsize{10}{14}\selectfont{
		\begin{tabular}{rll} \hline
			\textbf{Input}:  & (a) A local profile $z_a(x,y)$ before the film was deposited                                  & //Tab. \ref{tab3}               \\
			                 & (b)  A local profile $z_b(x,y)$  after the film was deposited                                 & //Tab. \ref{tab3}               \\
			\textbf{Output}: & Change of curvature tensor $\Delta\kappa$ in the primary coordinate system                    & //(\ref{Eq31})                  \\ \hline \hline
			\textbf{Begin}   &                                                                                               &                                 \\
			1.               & Find the difference $\Delta{z}$ of wafer surface profiles;                                    & //(\ref{Eq36})                  \\
			2.               & Identify the quadric $\Delta{z}=f(x,y)$;                                                      & //Tab. \ref{findQuadric}        \\
			3.               & Find the above quadric in the principal axes of its quadratic form;                           & //Tab. \ref{quadricInPrincipal} \\
			4.               & At point $p_0$, find principal components of the curvature tensor change:                     & //(\ref{Eq36})                  \\
			                 & (a) Estimate $\Delta\kappa{'}_{11}$ as the second derivative of (35) with respect to $x^{'}$; &                                 \\
			                 & (b) Estimate $\Delta\kappa{'}_{22}$ as the second derivative of (35) with respect to $y^{'}$; &                                 \\
			5.               & Estimate the tensor  $\Delta\kappa$ in the primary coordinate system;                         & //(\ref{Eq31})                  \\
			\textbf{End}.    &                                                                                               &                                 \\ \hline
		\end{tabular}}
\end{table}
\subsection{The third simplification - algorithms B31 and B32}
The last modification of the algorithm is the most radical one. In particular, this modification does not require finding the coordinate system associated with the principal axes of the quadratic form of the previously identified quadric. This means that it does not require solving the eigenproblem for the matrix of this quadratic form.

There are two versions of the algorithm.
The difference between them results from the way in which the quadric $f(x,y)$ is defined at point $p_0$, which describes the change in the wafer profile resulting from the deposition of a thin film.
If the quadric is given by the formula (\ref{Eq35}), the proposed version of the algorithm will be labeled as B31.
In the case when the quadric is given in the formula (\ref{Eq37}), the proposed version of the algorithm will be labeled as B32.
For a given version of the $f(x,y)$ quadic, the change of the curvature tensor of this profile, at the point $p_0$ can be described by the Hessian matrix (or Hessian):
\begin{equation}\label{Eq39}
	\Delta\kappa = \begin{bmatrix}
		\frac{\partial{^{2}f}}{\partial{x^2}}          & \frac{\partial{^{2}f}}{\partial{x}\partial{y}} \\
		\frac{\partial{^{2}f}}{\partial{y}\partial{x}} & \frac{\partial{^{2}f}}{\partial{y^2}}
	\end{bmatrix}.
\end{equation}
Because partial derivatives of quadric are calculated in expression (\ref{Eq39}), they can be expressed by appropriate coefficients of the quadric (\ref{Eq07}):
\begin{equation}\label{Eq40}
	\Delta\kappa = \begin{bmatrix}
		2a & c  \\
		c  & 2b
	\end{bmatrix}.
\end{equation}
A detailed version of this algorithm is shown in Table \ref{AlgorithmB3}.
\begin{table}
	\centering
	\caption{B31 and B32 algorithms - The third simplification of the algorithm to identify the change in the tensor of curvature - it uses the analytical properties of the quadric used to model the surface profile of the wafer}\label{AlgorithmB3}
	\fontsize{10}{14}\selectfont{
		\begin{tabular}{rll} \hline
			\textbf{Input}:  & (a) A local profile $z_a(x,y)$ before the film was deposited                       & //Tab. \ref{tab3}              \\
			                 & (b)  A local profile $z_b(x,y)$  after the film was deposited                      & //Tab. \ref{tab3}              \\
			\textbf{Output}: & Change of curvature tensor $\Delta\kappa$ in the primary coordinate system         & //(\ref{Eq31})                 \\ \hline \hline
			\textbf{Begin}   &                                                                                    &                                \\
			1.               & Identify quadric describing difference in wafer surface profiles;                  & //(\ref{Eq35}) or (\ref{Eq38}) \\
			2.               & Estimate the tensor $\Delta\kappa$ as a Hessian (\ref{Eq39}) of the given quadric; & //(\ref{Eq40})                 \\
			\textbf{End}.    &                                                                                    &                                \\ \hline
		\end{tabular}}
\end{table}
\section{Examples of the use of algorithms}
The article proposes five  versions of the two-dimensional Stoney algorithm. With regard to the first version of the algorithm, the remaining four are simplified to a different degree. Therefore, it can be expected that all these versions will give slightly different results. To compare the accuracy of these algorithms, they were all implemented to compare their accuracy.

\begin{figure}
	\centering
	{\subfloat[] {\label{mapaFull}
			\includegraphics[width=0.47\textwidth]{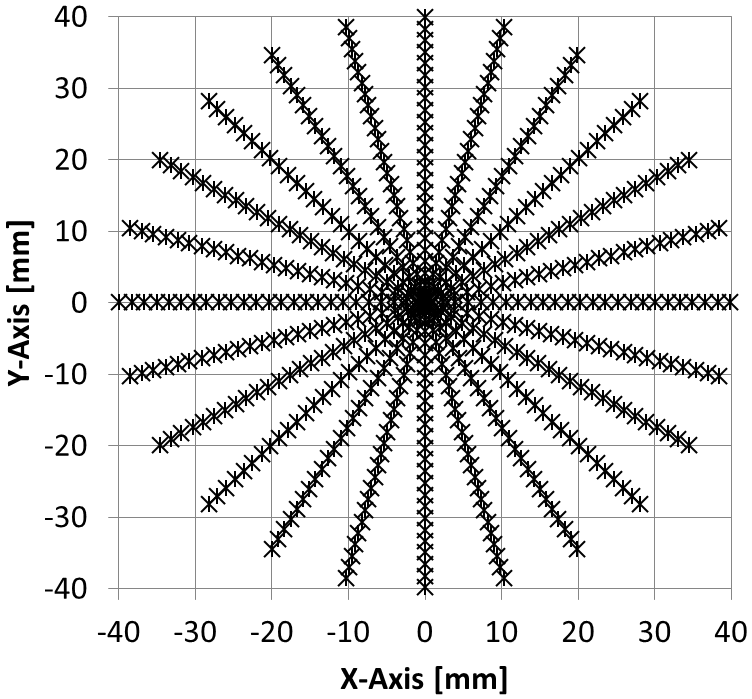}
		}
		\quad
		\subfloat[] {\label{areas}
			\includegraphics[width=0.47\textwidth]{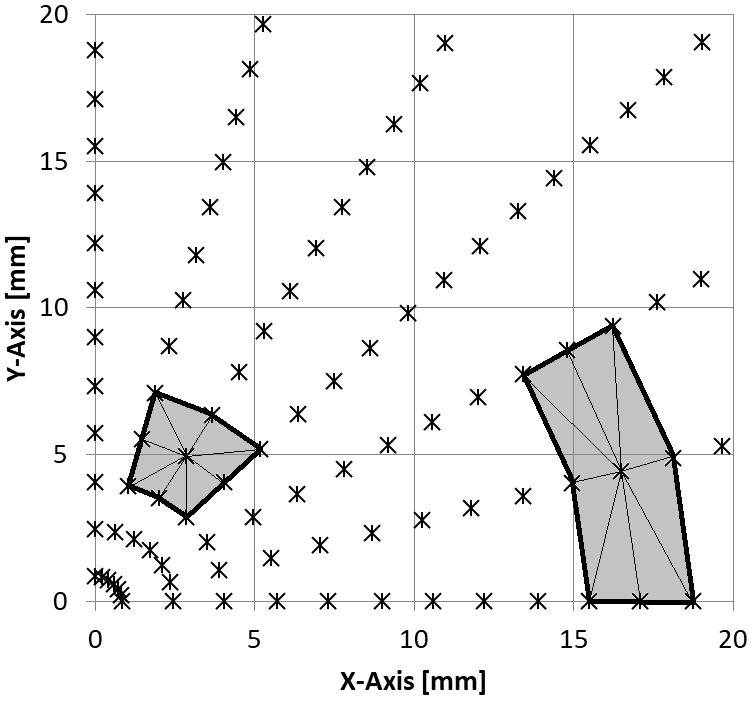}
		}
		\caption{Distribution of measuring points of wafer surface profile that were used to identify the curvature tensor: (a) map of measurement points; (b) examples of areas that were used to identify the tensor of curvature at a given point.}\label{mapa}}
\end{figure}

The results of wafer measurements with a diameter of 4 inches were used for the study.
The measurement was made 12 times along the diameter of the wafer, changing the measurement direction every 15 degrees.
Figure \ref{mapa} shows the map of the measured points, as well as the method for selecting points to identify the quadrics.
Figure \ref{areas}, which is a fragment of the map shown in Figure \ref{mapaFull}, presents two examples of central points $p_0$, as well as their neighbors, used to identify the quadrics needed to estimate curvature tensors.
The areas in Figure \ref{areas} in gray, including the center point and its nearest neighbors, show the area around which the quadric needed to estimate the tensor of curvature at this point will be identified.
Detailed results of wafer surface measurements are presented in Figure \ref{profile}.
Figure \ref{before} shows the results of measurements of the wafer surface profile before depositing a thin film.
Figure \ref{after} shows the results of measurements of the wafer surface profile after deposition of a thin film.
Figure \ref{differ} shows the difference between the two wafer surface profiles.
This difference shows how the surface profile of the wafer has changed as a result of stresses embedded in the technological process.
\begin{figure}
	\centering
	{\subfloat[] {\label{before}
			\includegraphics[width=0.47\textwidth]{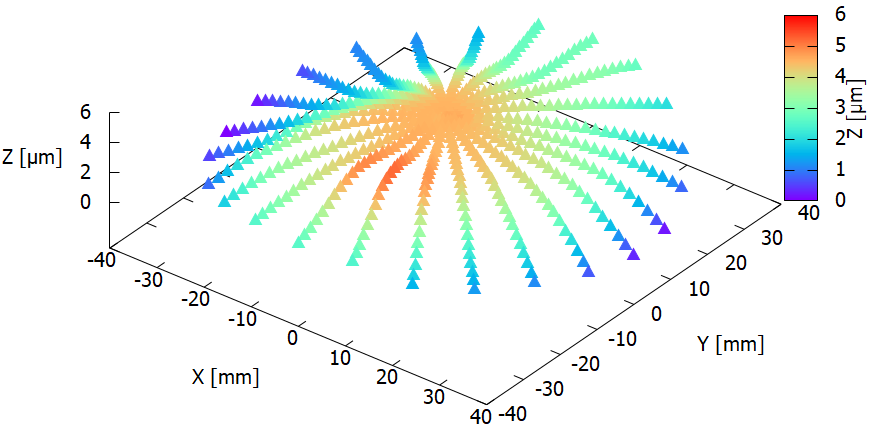}
		}
		\quad
		\subfloat[] {\label{after}
			\includegraphics[width=0.47\textwidth]{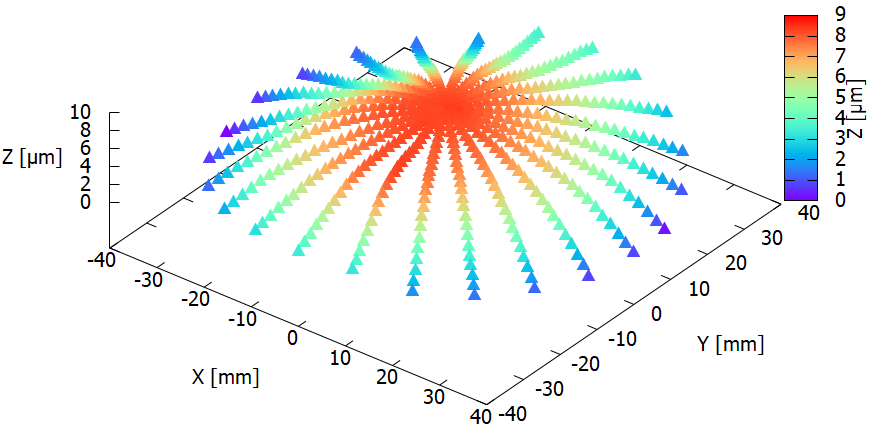}
		}
		\quad
		\subfloat[] {\label{differ}
			\includegraphics[width=0.47\textwidth]{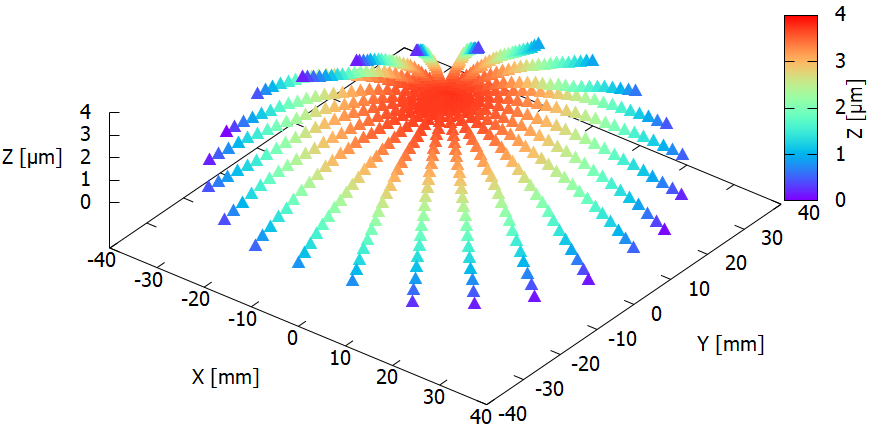}
		}
		\caption{The results of measurements of wafer surface profile: (a) before depositing a thin film; (b) after depositing a thin film; (c) the difference between these two profiles.}\label{profile}}
\end{figure}
\subsection{Algorithm A - no simplifications}
It was assumed that in the small neighborhood of a given point $p_0$, the surface profile of the wafer can be well described by means of a second degree surface, called a quadric. Having identified the quadric, which at the given point describes the surface profile of the wafer, you can find the curvature of this profile at this point. The analyzed algorithm (labeled in this article as Algorithm A) did not assume any other simplifications.

Change of curvature tensor was identified locally at each point, based on the measurements of both wafer surface profiles: first the profile measured before the film was deposited, and then the profile measured after the film was deposited.
Using the algorithm presented in section \ref{Subsection:localCurvature}, distributions of the tensor components of the change of curvature $\Delta\kappa$ were estimated.
There are several possibilities to present the results of the estimated curvature:
\begin{itemize}
	\item  The tensor components can be represented in a coordinate system whose axes are parallel to the axes of a given global coordinate system. Then, the graphs should show all three components of the change of curvature tensor $\Delta\kappa_{11}$, $\Delta\kappa_{22}$ as well as $\Delta\kappa_{12}$.
	\item  The change of curvature  can be represented so that each of its components is locally referenced to the coordinate system, in which one axis of the system is directed along the wafer's radius, while the other axis will be perpendicular to it. It is also necessary to show all three components  $\Delta\kappa_{11}$, $\Delta\kappa_{22}$ and $\Delta\kappa_{12}$ of the change of curvature  in the graphs.
	\item  	The components of the change of curvature  tensor  at a given point can be presented in the coordinate system of the principal axes of this  change of curvature  tensor.
\end{itemize}
The third method has some advantages over the first and second methods. Because in the coordinate system of the principal axes, the $\kappa_{12}$ shear component of the tensor disappears, and stress is proportional to the change of curvature tensor, therefore the future interpretation of stress will be simpler: at the given point of the space there is only a normal stretching as well as normal compression.
The advantage of this method of presenting the change of curvature tensor is also the fact that the diagrams show the distributions of two non-zero components of the change of curvature tensor, not the three components of the change of curvature tensor, as in the previous cases.

However, it should be noted that the third way of presenting the change of curvature tensor also has a disadvantage. The local coordinate systems of the principal axes, at different points, can be mutually rotated relative to each other, and this can not be clearly shown on the graph.

According to the third proposition, in this article in Figure \ref{distribOfKappa} the maps of the distribution of change of curvature  in the coordinate system  of its principal axes are presented.
Figures \ref{kappa11} and \ref{kappa22} show distributions of non-zero components of the $\Delta\kappa$ change of curvature tensor, respectively of the normal component $\Delta\kappa_{11}$ and the normal component $\Delta\kappa_{22}$.
\begin{figure}
	\centering
	{\subfloat[] {\label{kappa11}
			\includegraphics[width=0.47\textwidth]{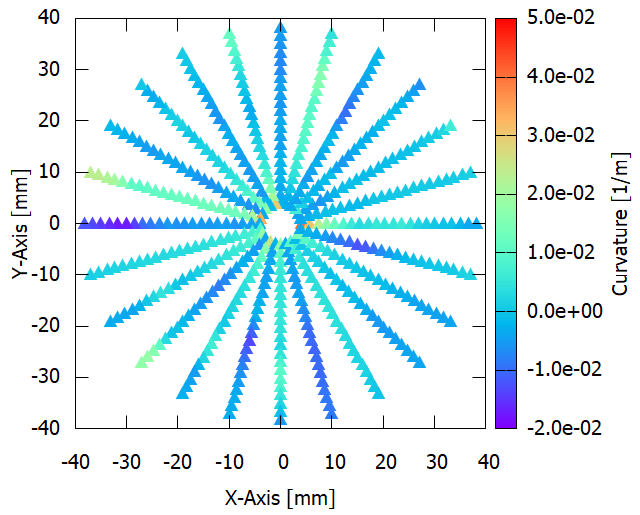}
		}
		\quad
		\subfloat[] {\label{kappa22}
			\includegraphics[width=0.47\textwidth]{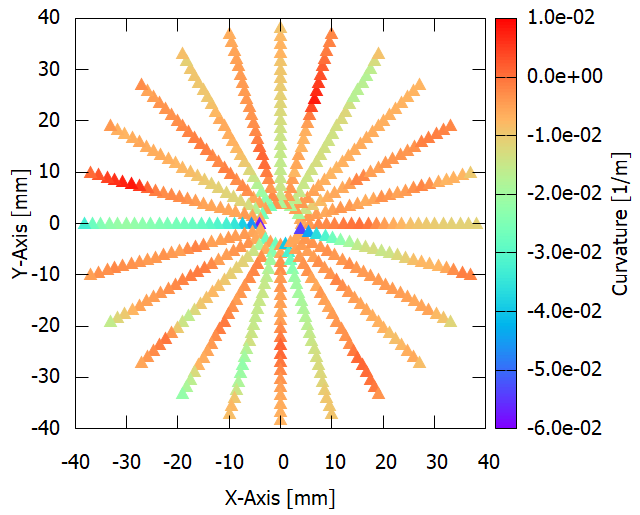}
		}
		\caption{Distributions of principal components of the change of curvature tensor: (a)  normal component $\Delta\kappa_{11}$; (b) normal component  $\Delta\kappa_{22}$.}\label{distribOfKappa}}
\end{figure}
A quick analysis of the results obtained shows that the normal component $\Delta\kappa_{11}$ of the curvature is dominated by negative values, while the normal component $\Delta\kappa_{22}$ contains mainly positive values. This means that in the plane of the wafer, in the coordinate system formed by the principal axes of stress, the deposited layer of thin film will be compressed along one axis, and stretched along the perpendicular to it the second axis.

On the other hand, a more accurate color analysis in Figure \ref{distribOfKappa} shows that for a fixed angle, the colors along the radius change continuously.
Unfortunately, the situation is worse at a fixed distance from the center of the wafer. In the distribution of the component of changes in curvature as a function of the angle, quite rapid color changes can be noticed.
This indicates relatively rapid transitions from low to high values and vice versa.
\begin{figure}
	\centering
	{\subfloat[] {\label{vs_Radius}
			\includegraphics[width=0.47\textwidth]{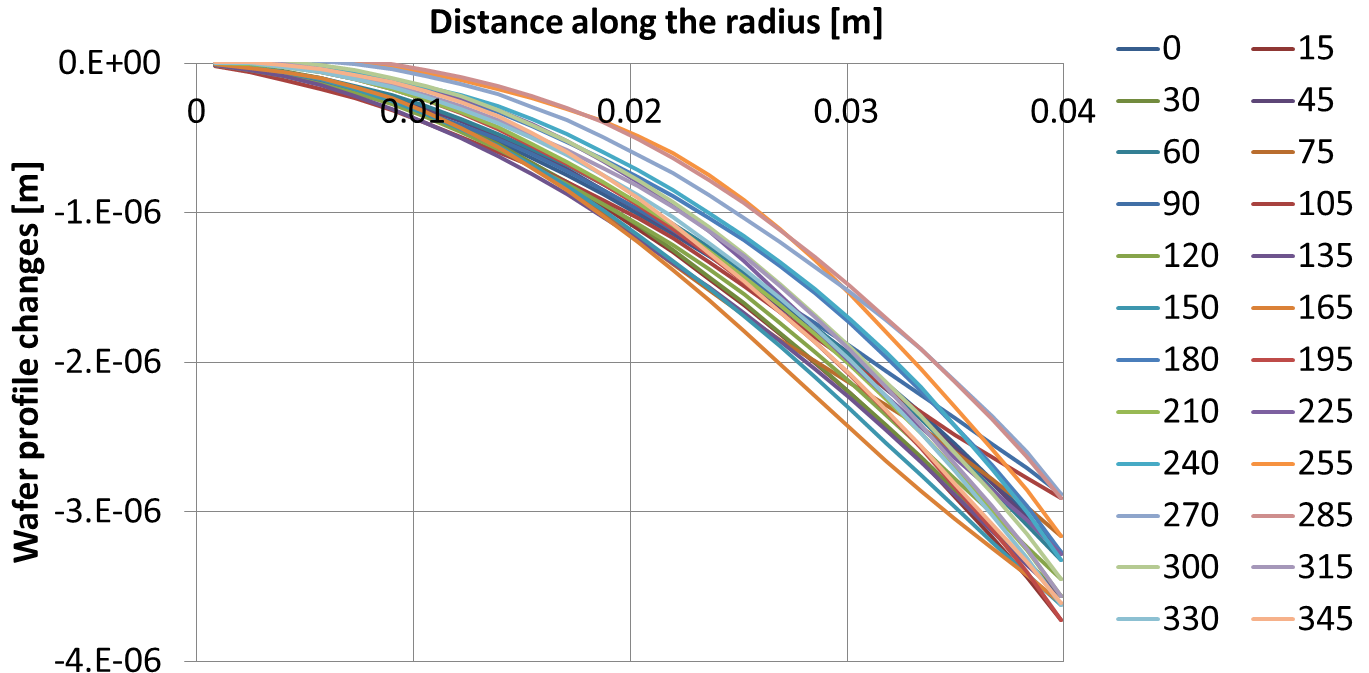}
		}
		\quad
		\subfloat[] {\label{vs_Angle}
			\includegraphics[width=0.47\textwidth]{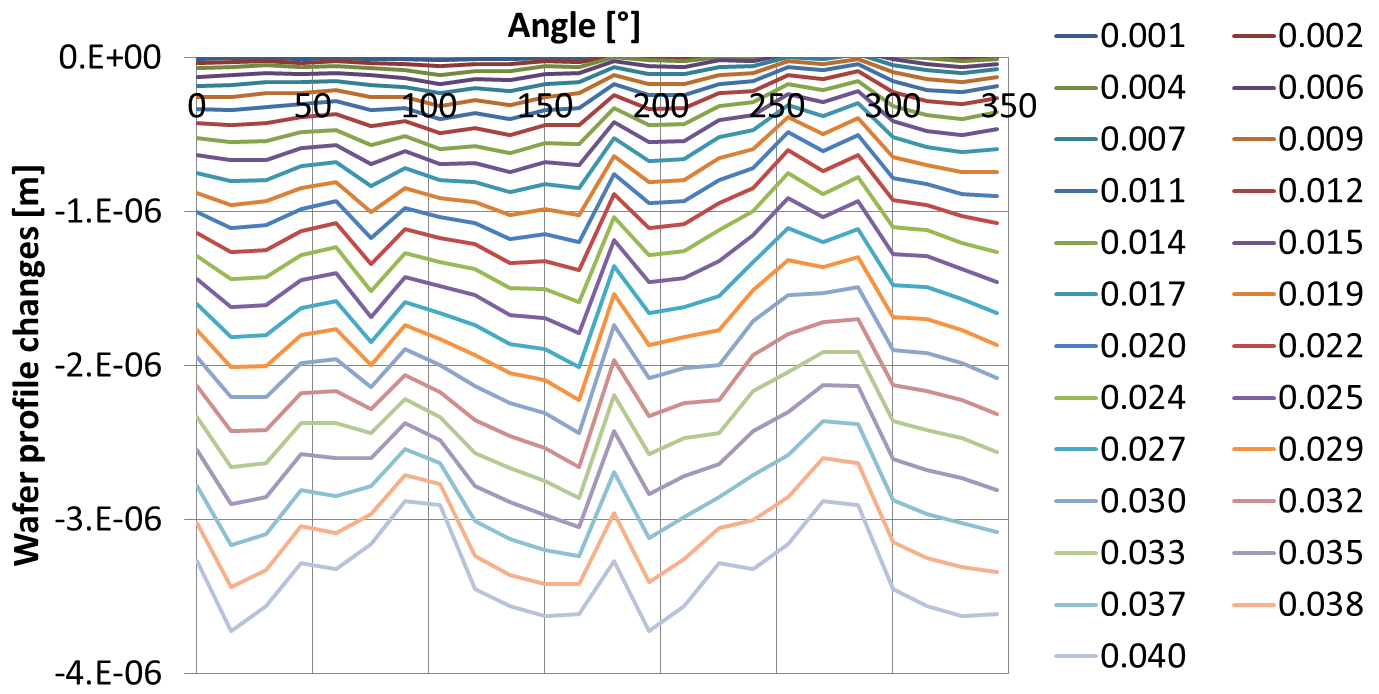}
		}
		\caption{Changes in the surface profile of the wafer: a) as a function of the distance from the center of the wafer, with a constant angle of measurement (in degrees); b) as a function of the angle, at a fixed distance from the center of the wafer (in meters).\label{profileChangeDistrib}}}
\end{figure}
These observations are in accordance with the diagrams shown in Figure \ref{profileChangeDistrib}.
At a fixed angle, the wafer surface profile changes along the radius create a smooth function (Figure \ref{vs_Radius}).
On the other hand, at a fixed distance from the center of the wafer, the changes in the wafer surface profile as a function of the angle indicate a mutual shift (in the vertical direction) of individual measurements (Figure \ref{vs_Angle}).
Therefore, it is reasonable to hypothesize that wafer profile measurements were made with the intention of using them to estimate one-dimensional stress, based on the one-dimensional Stoney algorithm.
These measurements were to accurately reflect changes in the profile of the wafer along the measurement line, and did not take into account the need for a good mapping between neighboring measurement lines.

\subsection{Simplified versions of algorithms}
It has been noticed that if condition (\ref{Eq31a}) is satisfied in expression (\ref{Eq04}), the algorithm labeled as Algorithm A can be simplified.
In practice, this means that the $d$ and $e$ components in the formula (\ref{Eq07}) describing the quadric should take values much smaller than one.
While calculating the difference in tensors of curvature using algorithm A, the values of coefficients $d$ and $e$ were calculated as intermediate results at each point. Therefore, it was possible to analyze them.
The results of the analysis of basic statistics for both quantities are presented in Table \ref{valOfDerivatives}. It has been found that the required condition is met: the corresponding values of the derivatives are over $2300$ times smaller than one.
Therefore, it is possible to check how accurate the simplified versions of the curvature tensor estimation algorithm are, compared to an algorithm that does not use simplification.
Assuming that the reference algorithm is Algorithm A, it is possible to compare errors generated by simplified versions of algorithms B1 and B2.
It is also possible to compare the differences in the results between B1 or B2 algorithms, and their simpler successors, i.e. the B31 and B32 algorithms. As a result of this analysis, it is possible to propose recommendations regarding the possibility of using one of the simplified algorithms B instead of the more complex algorithm A.

\begin{table}
	\centering
	\caption{Basic statistics of estimated derivatives}\label{valOfDerivatives}
	\fontsize{9.5}{13.5}\selectfont{
		\begin{tabular}{c||c|c|c|c} \hline
			Statistics         & $d_a$     & $e_a$     & $d_b$     & $e_b$     \\ \hline \hline
			Average            & 3.08E-06  & -1.20E-05 & 3.29E-06  & -1.65E-05 \\ \hline
			Median             & -7.00E-06 & -2.56E-05 & -5.58E-06 & -3.85E-05 \\ \hline
			Standard deviation & 1.01E-04  & 5.61E-05  & 1.72E-04  & 1.12E-04  \\ \hline
			Minimum            & -2.74E-04 & -1.10E-04 & -4.31E-04 & -1.88E-04 \\ \hline
			Maximum            & 2.17E-04  & 1.94E-04  & 3.29E-04  & 3.13E-04  \\ \hline
		\end{tabular}}
\end{table}
\subsubsection{Comparison of the B1 algorithm with the A algorithm}
In this work, the accuracy of the B1 algorithm was also examined.
The results obtained using this algorithm were compared with the results obtained using the reference algorithm A.
As in the case of algorithm A, a tensor of curvature was identified at 552 points.
For both algorithms, the results obtained in the primary coordinate system were compared.
Table \ref{errorB1} shows the basic statistics of relative errors estimated for the components of the $\Delta\kappa$ curvature tensor.
For the $\Delta\kappa_{22}$ component, the relative error value does not exceed 0.008\%. For other components, the value of this error is even smaller.
It should be noted that the results obtained with the use of the B1 algorithm are promising.

\begin{table}
	\centering
	\caption{Relative error of results obtained using the B1 algorithm in comparison with the reference algorithm A}\label{errorB1}
	\fontsize{9.5}{13.5}\selectfont{
		\begin{tabular}{c||c|c|c} \hline
			Statistics         & $\epsilon_{\kappa_{11}}$ & $\epsilon_{\kappa_{22}}$ & $\epsilon_{\kappa_{12}}$ \\ \hline \hline
			Average            & -0.00002\%               & -0.00002\%               & 0.00001\%                \\ \hline
			Median             & 0\%                      & 0\%                      & 0\%                      \\ \hline
			Mode               & 0\%                      & 0\%                      & 0\%                      \\ \hline
			Standard deviation & 0.0002\%                 & 0.0004\%                 & 0.0003\%                 \\ \hline
			Minimum            & -0.0026\%                & -0.0078\%                & -0.0019\%                \\ \hline
			Maximum            & 0.0006\%                 & 0.0007\%                 & 0.0024\%                 \\ \hline
		\end{tabular}}
\end{table}

\subsubsection{Comparison of the B2 algorithm with the A algorithm}
As in the case of the B1 Algorithm, the accuracy of the B2 algorithm was also examined.
Here, similarly as before, the results obtained using this algorithm were compared to the results obtained using the reference algorithm A.
Table \ref{errorB2} shows the basic statistics of relative errors $\epsilon$ estimated for tensor $\Delta\kappa$ components, describing the difference in curvatures, calculated using the B2 algorithm. It should be noted that in this case larger errors were obtained.
In one case, the maximum relative error reaches almost 4.4\%.
Analyzing all cases in which curvature was estimated, it was found that only in five cases the relative error value exceeded 1\%.
If the possibility of a slightly larger error is allowed at the level of individual percentages, then this algorithm can be applied in practice. It is all the more justified that its undoubted advantage is greater simplicity than other algorithms.
\begin{table}
	\centering
	\caption{Relative error of results obtained using the B2 algorithm in comparison with the reference algorithm A}\label{errorB2}
	\fontsize{9.5}{13.5}\selectfont{
		\begin{tabular}{c||c|c|c} \hline
			Statistics         & $\epsilon_{\kappa_{11}}$ & $\epsilon_{\kappa_{22}}$ & $\epsilon_{\kappa_{12}}$ \\ \hline \hline
			Average            & -0.0002\%                & -0.0079\%                & 0.0104\%                 \\ \hline
			Median             & 0\%                      & 0\%                      & 0\%                      \\ \hline
			Mode               & 0\%                      & 0\%                      & 0\%                      \\ \hline
			Standard deviation & 0.02\%                   & 0.13\%                   & 0.22\%                   \\ \hline
			Minimum            & -0.33\%                  & -2.79\%                  & -1.56\%                  \\ \hline
			Maximum            & 0.16\%                   & 0.04\%                   & 4.37\%                   \\ \hline
		\end{tabular}}
\end{table}

\subsubsection{Comparison of algorithms B31 with B1, as well as B32 with B2}
The accuracy of algorithms B31 and B32 was also examined in the work.
It was noticed that the results obtained using the B31 algorithm are identical to the results obtained with the B1 algorithm.
Analogously, the results obtained with the B32 algorithm are identical to the results obtained with the B2 algorithm.
On the one hand, the algorithms B31 and B32 have the same accuracy as the B1 and B2 algorithms respectively.
On the other hand, algorithms B31 and B32 are simpler than the corresponding algorithms B1 and B2.
This means that if you decide not to use the A algorithm, you do not need to use the B1 and B2 algorithms.
Depending on the expected accuracy of calculations, it is enough to use the B31 algorithm or the B32 algorithm.
\section{Conclusions}
This article presents the extension of the one-dimensional Stoney algorithm to a two-dimensional case. The proposed extension involves the modification of the curvature estimation method without the simultaneous modification of other components calculated in the one-dimensional Stoney algorithm.
First, an algorithm was proposed to which the A label was assigned. Next, four additional versions of the two-dimensional Stoney algorithm were proposed, labeled as B1 and B2, as well as B31 and B32.
In order to propose a recommendation for an algorithm suitable for practical use, all proposed algorithms should be compared in the context of their complexity. For numerical algorithms (all proposed algorithms are just numerical algorithms) their accuracy should also be compared.

First of all, we must state that algorithm A is the most complex, but also the most accurate. Both its greatest complexity and the highest accuracy result from the fact that this algorithm does not use almost any simplifying assumptions.
The only simplifying assumption is the assumption about the possibility of modeling the wafer surface profile with the help of a second-degree surface called a quadric. Therefore, it was assumed that when evaluating other algorithms, this algorithm will be considered as a reference algorithm.

All versions of the algorithm of identification  of the change  of curvature  tensor were created in such a way that the reference algorithm A was simplified by adopting some additional assumptions.
It should also be noted here that each subsequent version of the algorithm was simpler than the version preceding it, but it was not necessarily less accurate from it.

\begin{table}
	\centering
	\caption{The relationship between the ranking of the complexity of algorithms and the ranking of the accuracy of these algorithms}\label{compleXvsAccur}
	\fontsize{10}{14}\selectfont{
		\begin{tabular}{c|c|c|c|c} \hline
			\multicolumn{2}{c|}{\ } & \multicolumn{3}{|c}{Complexity}                 \\ \cline{3-5}
			\multicolumn{2}{c|}{\ } & $1$                             & $2$ & $3$     \\ \hline
			                        & $1$                             &     &     & A \\ \cline{2-5}
			Accuracy                & $2$                             &     & B31 &   \\ \cline{2-5}
			                        & $3$                             & B32 &     &   \\ \hline
		\end{tabular}}
\end{table}

To assess the complexity of algorithms, in its structure, some operations are distinguished that can be considered as dominant operations. In the discussed algorithms, two operations can be considered dominant. The first of these is the identification of a quadric, which requires the creation of a matrix of a system of 6x6 normal equations, and then the solution of this system. The second dominant operation is solving eigenproblem and coordinate system rotation, which operates on a 2x2 matrix. At least one of the two operations mentioned above was implemented in each of the algorithms discussed here.

During the tests, it was observed that the total time of operation of the program estimating changes in curvature at all measuring points, running on a standard personal computer, was lower than 0.01 seconds.
This means that from the point of view of the program's duration, its complexity is not a critical issue.
The complexity of the algorithm is important from the point of view of the process of creating a computational program that uses the algorithms discussed here: a simpler program can be built from simpler algorithms. From this point of view, it is also not indifferent to whether we require an algorithm to identify the quadric, and to solve the eigenproblem, or whether we require that he only be able to identify the quadric.

Not deciding whether the identification of the quadric is more complex than solving the eigenproblem and rotation of the coordinate system, or whether the reverse is true, the real ranking of the complexity of the algorithms takes the following form:
\begin{enumerate}
	\item B32 algorithm - requires one-time identification of the quadric;
	\item B2 and B31 algorithms:
	      \begin{itemize}
		      \item The B2 algorithm - requires one-time identification of the quadric and single-solution eigenproblem and rotation of the coordinate system;
		      \item B31 algorithm - requires double identification of the quadric;
	      \end{itemize}
	\item The B1 algorithm - requires double identification of the quadric and two-fold solution of eigenproblem and rotation of the coordinate system;
	\item Algorithm A - requires double identification of the quadric and two-fold solution of eigenproblem and rotation of the coordinate system. Unlike in previous cases, the method of calculating the curvature does not assume any simplifications.
\end{enumerate}

On the other hand, because the curvature estimation algorithms are numerical algorithms, then for each proposed algorithm, their mutual accuracy was tested. This accuracy, starting from the most-accurate reference algorithm A, is presented in the following ranking:
\begin{enumerate}
	\item Algorithm A;
	\item B1 and B31 algorithms;
	\item B2 and B32 algorithms.
\end{enumerate}
In the context of the analysis of the accuracy and computational complexity of the curvature estimation algorithms, two facts should be noted:
\begin{itemize}
	\item The B1 algorithm has the same accuracy as the B31 algorithm, whereas the B2 algorithm has the same accuracy as the B32 algorithm;
	\item The B1 algorithm has greater complexity than the B32 algorithm, while the B2 algorithm has greater complexity than the B32 algorithm.
\end{itemize}
Intuition suggests that if two algorithms that solve the same problem have identical accuracy, then considering the possibility of their implementation, the algorithm of greater complexity should be omitted.
Therefore, algorithms B1 and B2 will be omitted, and only algorithms A, B31 and B32 will be considered.
Now the rankings of complexity and accuracy for these three remaining algorithms can be presented in Table \ref{compleXvsAccur}. In this table you can see what are the relationships between the complexity of algorithms and their accuracy.

The question remains for recommendations for one of the algorithms proposed in the article. The author of this article thinks that the B32 algorithm should be recommended for practical use.
On the one hand, it is an algorithm absolutely the simplest of all algorithms discussed in this article.
On the other hand, its accuracy, although it is smaller than the accuracy of other algorithms, is not a disqualifying accuracy.
In five cases, the error obtained with this algorithm was at a level not exceeding 5\%. In the remaining, much more numerous cases, the error of curvature estimation was less than 1\%.
In addition, there are some symptoms (Figure \ref{profileChangeDistrib}), which indicate that the main source of errors in curvature estimation may be errors in the wafer surface profile measurement, and not the algorithm calculating the curvature.
Considering the above-mentioned circumstances, the recommendation for the B32 algorithm is unquestionable.

\section*{Acknowledgments}
The measurements were made with the {\textregistered}Tencor FLX-2320 device. The results of the measurements were obtained courtesy of Dr. Marek Guziewicz from the Institute of Electron Technology in Warsaw.

\bibliography{stoney2Dfull}\label{bibliography}
\bibliographystyle{IEEEtran}
\end{document}